\documentclass[letterpaper, 10 pt, conference]{ieeeconf}  

\IEEEoverridecommandlockouts                              

\usepackage{balance} 
\usepackage{booktabs}
\usepackage{multicol}
\usepackage{multirow}
\usepackage{graphicx}
\usepackage{algorithm}
\usepackage[noend]{algpseudocode}
\usepackage{tcolorbox}
\usepackage{amsmath}

\title{\LARGE \bf
LLMDR: LLM-Driven Deadlock Detection and Resolution in Multi-Agent Pathfinding
\thanks{This work has been submitted to the IEEE for possible publication. Copyright may be transferred without notice, after which this version may no longer be accessible.}
}

\author{Seungbae Seo$^{1}$, Junghwan Kim$^{2}$, Minjeong Shin$^{3}$, and Bongwon Suh$^{4}$
\thanks{$^{2,3,4}$Junghwan Kim, Minjeong Shin, Bongwon Suh are with Seoul National University, 1 Gwanak-ro, Gwanak-gu, Seoul, Republic of Korea, 08826.}%
\thanks{$^{1}$Seungbae Seo: {\tt\small wiki1177@snu.ac.kr}}%
\thanks{$^{2}$Junghwan Kim: {\tt\small jhbale11@snu.ac.kr}}%
\thanks{$^{3}$Minjeong Shin: {\tt\small shinmj1024@snu.ac.kr}}%
\thanks{$^{4}$Bongwon Suh: {\tt\small bongwon@snu.ac.kr}}%
}

\begin{document}

\maketitle
\thispagestyle{empty}
\pagestyle{empty}

\begin{abstract}

Multi-Agent Pathfinding (MAPF) is a core challenge in multi-agent systems. Existing learning-based MAPF methods often struggle with scalability, particularly when addressing complex scenarios that are prone to deadlocks. To address these challenges, we introduce LLMDR (LLM-Driven Deadlock Detection and Resolution), an approach designed to resolve deadlocks and improve the performance of learned MAPF models. LLMDR integrates the inference capabilities of large language models (LLMs) with learned MAPF models and prioritized planning, enabling it to detect deadlocks and provide customized resolution strategies. We evaluate LLMDR on standard MAPF benchmark maps with varying agent numbers, measuring its performance when combined with several base models. The results demonstrate that LLMDR improves the performance of learned MAPF models, particularly in deadlock-prone scenarios, with notable improvements in success rates. These findings show the potential of integrating LLMs to improve the scalability of learning-based MAPF methods.

\end{abstract}


\section{INTRODUCTION}

\begin{figure}
    \centering
    \includegraphics[width=1\linewidth]{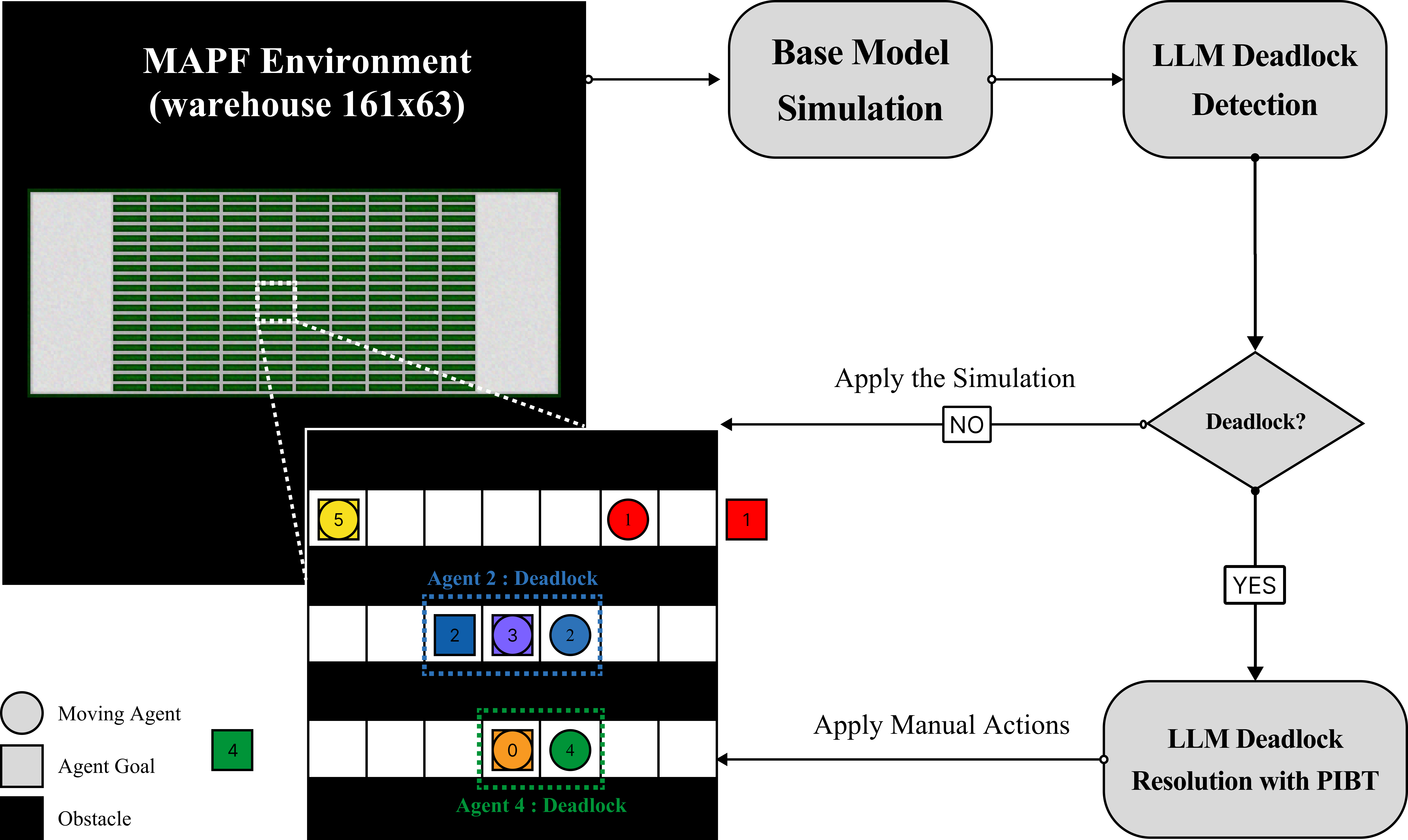}
    \caption{Overview of LLMDR (LLM-Driven Deadlock Detection and Resolution). When a deadlock is detected in the base model simulation, manual actions are iterated to resolve the deadlock. If no deadlock is detected, the base model simulation is directly applied to the environment.}
    \label{fig:overview}
\end{figure}

Multi-Agent Pathfinding (MAPF) \cite{stern2019multi} is a core challenge in multi-agent systems. Its applications span various domains, such as autonomous vehicles \cite{li2023intersection}, multi-robot systems \cite{kaduri2020algorithm}, and video games \cite{ma2017feasibility}. The objective of MAPF is to coordinate multiple agents to find collision-free paths to their individual goals while minimizing the overall number of steps. As the number of agents and the complexity of environments increase, solving MAPF instances becomes increasingly difficult \cite{kraus1997negotiation}.

Existing approaches to MAPF can be broadly classified into algorithmic and learning-based methods. Algorithmic methods include systematic search algorithms like Conflict-Based Search (CBS) \cite{boyarski2015icbs}, which struggle with large-scale scenarios due to their exponential worst-case time complexity. Recently, learning-based methods have shown promise in scaling across different scenarios. For instance, PRIMAL \cite{sartoretti2019primal} combines reinforcement learning with imitation learning, while more recent approaches like DHC \cite{ma2021dhc} and DCC \cite{ma2021dcc} have introduced advanced communication mechanisms between agents. However, learning-based methods still face challenges in behavior consistency and low-level optimality, especially in complex scenarios. Failures in MAPF policies trained through reinforcement learning are not always easy to identify \cite{ma2021dhc, sartoretti2019primal}. Moreover, deadlock situations often involve multiple agents, making simple or random resolution strategies ineffective \cite{flammini2024preventing}.

To address these challenges, we present \textbf{LLMDR (LLM-Driven Deadlock Detection and Resolution)}, which integrates large language models (LLMs) with learned MAPF models and prioritized planning. LLMDR has two phases: deadlock detection by LLM and resolution using LLM strategies with Priority Inheritance with Backtracking (PIBT) \cite{okumura2022priority}. In detection, the LLM checks the agents' states within a detection window to spot deadlocks. When deadlock is detected, the system shifts to resolution, where the LLM sets priorities for agents and actions. These priorities guide PIBT to generate collision-free 1-step moves. We tested LLMDR on standard MAPF benchmark maps \cite{stern2019multi} with 4 to 64 agents. The results show that LLMDR improves learned MAPF models, especially in complex environments. This suggests that many learned MAPF models fail due to deadlocks, and resolving them with LLMDR improves performance. LLMDR’s impact grows with more agents and increasing complexity, showing its potential to improve various learned MAPF models. \\

The main contributions of our work are:
\begin{itemize}
\item We present LLMDR, which combines large language models with learning-based MAPF methods and prioritized planning to detect and resolve deadlocks.
\item Through experiments, we show that LLMDR enhances various learned MAPF models, improving performance across environments and agent numbers, thus boosting the scalability of learning-based MAPF methods.
\end{itemize}


\section{Related Work}

In this section, we review learning-based MAPF methods and their limitations. Then, we explore recent developments in utilizing LLMs for planning tasks.

\subsection{Learning-Based MAPF Methods}
Learning-based MAPF methods using reinforcement learning (RL) to train policies for agents have been studied to improve generalization across diverse environments. PRIMAL \cite{sartoretti2019primal} combines RL with imitation learning from a centralized MAPF planner to train decentralized policies \cite{ferner2013odrm}. Subsequent studies have introduced the use of the A* algorithm as a key component \cite{grenouilleau2019multi, liu2020mapper, wang2020mobile}. Recent studies have facilitated communication between agents. The use of graph neural networks allowed agents to communicate \cite{li2020graph}. In DHC \cite{ma2021dhc}, pathfinding and communication strategies were trained separately, with agents communicating through a graph convolution mechanism. To conserve resources, DCC \cite{ma2021dcc} introduced decision-causal communication, which selects only relevant neighbors. SACHA \cite{lin2023sacha}, based on Soft Actor-Critic, trained the actor and critic network architectures.

Some studies have shifted focus towards resolving issues such as conflicts and deadlocks in learning-based methods. PIBT \cite{okumura2022priority} and LaCAM \cite{okumura2023lacam} have been combined with the action probability distributions generated by a learned model to prevent collisions \cite{veerapaneni2024improving}. EPH \cite{tang2024ensembling} employs several inference strategies and ensemble methods. RDE \cite{gao2023rde} proposed a method that detects deadlock based on specific rules by analyzing agents’ past actions. When a deadlock is detected, agents resolve it by randomly selecting an action from possible options. However, failures in learned MAPF models are not always easy to identify. In many cases, deadlock involves multiple agents, and therefore, it is important to account for the interdependencies between agents. For escape policies, strategies that adapt to the agents' states may be more effective than purely random methods.

\subsection{LLM for Planning}

Since LLMs can perform zero-shot planning and demonstrate robustness in novel environments \cite{huang2022language}, there have been attempts to incorporate them into multi-robot planning. Some studies have explored using LLMs to decompose high-level task commands and distribute them among multiple robots, facilitating coalition formation and task allocation \cite{kannan2023smart, shukla2023lgts}.

Efforts have also been made to apply LLMs in other fields where LLMs handle high-level control, while low-level control is managed through RL \cite{zhu2023ghost}. For example, the common deadlock problem in multi-robot systems (MRS) can be mitigated using a framework that combines LLM-driven high-level control with graph neural network-based low-level control \cite{garg2024large}.

In the context of MAPF, Chen et al. \cite{chen2024solving} investigated the direct use of LLMs. While successful in simple scenarios, their approach faced significant challenges in complex environments. This study highlighted LLMs' limitations in handling multi-agent coordination and spatial reasoning, particularly in complex scenarios. In contrast, LLMDR addresses these limitations by using LLMs for high-level tasks like deadlock detection and resolution, while relying on learned MAPF models for low-level planning. This approach allows LLMDR to benefit from LLMs' reasoning capabilities in decision-making, without being constrained by their limitations in handling low-level planning.


\section{Preliminaries}

This paper uses prioritized planning to enable cooperative behaviors based on LLM strategies to resolve deadlocks. While prioritized planning helps cooperation, it can cause agents to get stuck. To mitigate this, we apply PIBT \cite{okumura2022priority}, which improves prioritized planning by resolving conflicts.

\subsection{Multi-Agent Pathfinding}

The standard MAPF \cite{stern2019multi} problem is defined as follows: Given a set of agents \( A = \{1, 2, \dots, k\} \) on an undirected graph \( G = (V, E) \), where \( V \) represents locations and \( E \) denotes connections, the goal is to find collision-free paths from each agent’s start \( s(i) \in V \) to its target \( t(i) \in V \). Agents move in discrete time steps, either waiting or moving to an adjacent node. A move from \( v \) to \( v' \) is \( a(v) = v' \), while staying is \( a(v) = v \). An agent’s plan is \( \pi_i = (a_1, a_2, \dots, a_n) \), guiding it from \( s(i) \) to \( t(i) \). A valid solution provides \( k \) such plans \( \{\pi_1, \pi_2, \dots, \pi_k\} \), ensuring all agents reach their targets without collisions.

\begin{algorithm}
\scriptsize
\caption{\textbf{PIBT algorithm}}
\label{alg:PIBT}
\raggedright
\textbf{Parameters:} Current states $s^{1:N}$, actions $a^{1:5}_{1:N}$, Static Obstacles $W$ \\
\textbf{Output:} Collision free actions $a^{1:N}$
\begin{algorithmic}[1]
\Procedure{\textbf{PIBT}}{Current states $s^{1:N}$, \text{agentPriorities}}
    \State Reserved $\gets \emptyset$
    \State Moves $\gets \text{dictionary()}$
    \For{$k \in \text{argsort}(\text{agentPriorities})$}
        \If{$k \notin \text{Moves.keys()}$}
            \State \textbf{PIBT-H}($k$, $a_{1:5}^{1:N}$, $s^{1:N}$, Reserved, Moves)
        \EndIf
    \EndFor
    \State \Return Moves
\EndProcedure
\\
\Procedure{\textbf{PIBT-H}}{Agent $k$, Action $a_{1:5}^{1:N}$, Current States $s^{1:N}$, Reserved, Moves}
    \For{$a \in a_{1:5}^{k}$} \Comment{Sort $a_{1:5}^{k}$ with preferred $a$ first}
        \State $s' \gets T(s^k, a)$
        \If{$s' \in W \cup \text{Reserved}$ \textbf{or} $(s', s) \in \text{Reserved}$}
            \State \textbf{continue}
        \EndIf
        \State \text{Moves}[$k$] $\gets a$, \text{Reserved} $\gets$ \text{Reserved} $\cup \{s, (s, s')\}$
        \If{$\exists$ agent $j \neq k$ at $s'$}
            \If{\textbf{PIBT-H}($j$, $a_{1:5}^{1:N}$, $s^{1:N}$, Reserved, Moves)}
                \State \Return Success
            \EndIf
            \State \text{Moves}[$k$] $\gets \emptyset$, \text{Reserved} $\gets \text{Reserved} \setminus \{s, (s, s')\}$
        \Else
            \State \Return Success
        \EndIf
    \EndFor
    \State \Return Failure
\EndProcedure
\end{algorithmic}
\end{algorithm}

\subsection{Prioritized Planning and PIBT}

In MAPF, prioritized planning \cite{Erdmann-1987-15683} is used to coordinate agents by allowing higher-priority agents to act first. It promotes cooperative behaviors which helps prevent conflicts. However, while effective, prioritized planning can lead to issues like agents becoming stuck, particularly in complex environments.

To overcome these limitations, we incorporate PIBT algorithm \cite{okumura2022priority}, which extends prioritized planning by introducing mechanisms to resolve conflicts and prevent agents from getting stuck. PIBT allows lower-priority agents to temporarily inherit the priority of higher-priority agents if lower-priority agents are blocking the paths of the higher-priority agents. Additionally, PIBT employs backtracking to further enhance the algorithm's ability to manage conflicts. Algorithm \ref{alg:PIBT} outlines the execution of PIBT, where the priorities between agents and the priorities among each agent's actions are both considered when determining 1-step plans \cite{veerapaneni2024improving}. PIBT is simple to implement and computationally efficient, making it suitable for large-scale, real-time applications \cite{okumura2022priority}.

However, PIBT operates greedily, which can result in suboptimal solutions over the long term. The algorithm prioritizes resolving immediate conflicts between agents, potentially neglecting the global context of the problem. This can lead to a degradation in overall solution quality. In our approach, the priorities of agents and their actions are determined by strategies generated by LLMs. These priorities are then integrated into PIBT, which uses them to resolve conflicts for a single step. This process enables more effective deadlock management across multiple steps, contributing to improved long-term performance.


\section{LLMDR: LLM-Driven Deadlock Detection and Resolution}

\begin{tcolorbox}[colback=gray!10!white, colframe=black, title=Prompt 1: Prompt for Deadlock Detection]\label{prompt:deadlock_detection}
\scriptsize
You are given \{detection\_window\_length\} action logs of agents to detect deadlocks.

Follow these steps in order: \\

\textbf{1. Classify deadlocks}:
\begin{itemize}
\item Detect agents that are exhibiting deadlock conditions.
\item Deadlock conditions: No movement, Wandering

\item Not deadlocks: Always “Arrived", Arrived and stationary, Consistent movement

\end{itemize}

\textbf{2. Group deadlocked agents}:
\begin{itemize}
\item Group deadlocked agents that are within a 2-Manhattan distance of each other.
\item If a deadlocked agent is within a 2-Manhattan distance of an already arrived agent, include them in the same group.
\end{itemize}

\textbf{3. Provide solutions}:
\begin{itemize}
\item Use the “leader" method for independently deadlocked agents or when any agent in the group has a goal more than 8 Manhattan units away.
\item Use the “radiation" method when all agents in the group are near their goals (less than 8 units) and likely to experience repeated deadlocks.
\end{itemize}
\vspace{\baselineskip}
Below are the \{detection\_window\_length\} action logs of agents.\\

\{detection\_window\}\\

Provide the agent group status in this JSON format:\\
\{ 
“agent\_id": [Agent IDs in the same group],
“solution": “leader" or “radiation"
\}
\end{tcolorbox}

\begin{figure*}
    \centering
    \includegraphics[width=0.8\linewidth]{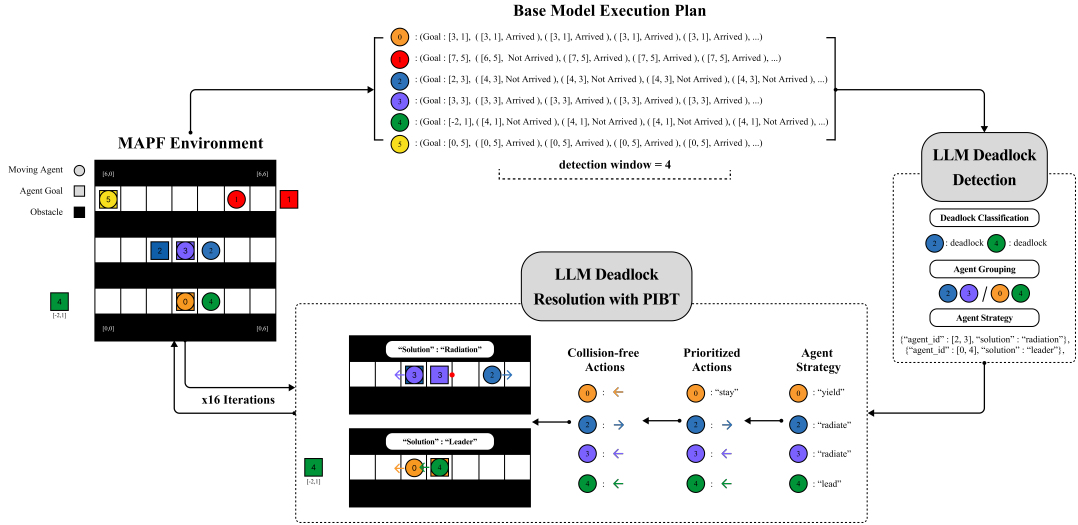}
    \caption{Detailed implementation of LLMDR (LLM-Driven Deadlock Detection and Resolution). The figure illustrates the process where agents in deadlock situations are detected by the LLM, and the action sets for each agent to resolve the deadlocks are obtained using the strategies proposed by the LLM along with PIBT algorithm. PIBT is used to derive collision-free actions from strategies generated by the LLM, thus enabling cooperative behavior among agents to resolve deadlocks.}
    \label{fig:method}
\end{figure*}

We introduce LLMDR, a novel approach to improve learned MAPF models. LLMDR operates in two main phases: deadlock detection by LLM, and deadlock resolution using LLM strategies with PIBT. Figure \ref{fig:method} shows the overall flow of LLMDR. In the following subsections, we will explain LLMDR in the order depicted in Figure \ref{fig:method}.

\subsection{Deadlock Detection by LLM}

The LLM detects potential deadlocks by analyzing agent movement within a specified detection window, extracted from the execution plan. The execution plan, generated by simulating the environment using the base model, logs agent positions as coordinate data over time. Both the execution plan and the detection window have predefined lengths, with the detection window representing the initial part of the execution plan used for deadlock analysis.

For deadlock detection, the LLM considers agents that have never reached their goal during the execution plan. Additionally, agents within a 9×9 area around them are included in the analysis. This ensures only agents potentially causing deadlocks are inspected.

Each agent’s state within the detection window is defined by three elements: (1) the agent’s ID, (2) the coordinates of the agent’s goal, and (3) the agent’s position at each step. The detection window is attached to the LLM prompt for deadlock analysis. Prompt 1, a simplified version of the original prompt, provides the detection window to the LLM, which goes through three main stages to detect deadlocks and suggest resolution strategies for each deadlock. First, the LLM analyzes the agents' states and classifies them as either in a deadlock or a non-deadlock. Second, the agents identified as being in a deadlock and any nearby deadlocked/arrived agents are grouped together, forming groups to address the deadlock issue. Third, the LLM provides appropriate strategies based on the characteristics of each deadlock group. \\

\noindent\textbf{Classify Deadlocks}~~The LLM classifies whether an agent is in a deadlock based on its behavior—specifically, whether the agent exhibits no movement or is wandering without consistent goal-oriented movement. To prevent the misclassification, the LLM is also provided with criteria for identifying agents that are functioning normally. \\

\noindent\textbf{Group Deadlocked Agents}~~The LLM groups deadlocked agents that are in close proximity to each other. For example, if a deadlocked agent is within a 2-Manhattan distance of another deadlocked agent or an agent that has already arrived at its goal, they are grouped together. The reason for grouping agents this way is that deadlocks often occur due to interactions between agents, making it necessary to resolve the deadlock collectively.
\\

\noindent\textbf{Assign Strategies to Deadlock Groups}~~The LLM assigns resolution strategies to each deadlocked group based on the characteristics of the deadlock. The “leader" method is applied to groups where at least one agent is significantly farther from its goal. Conversely, the “radiation" method is used for groups where all agents are near their goals, meaning they are at risk of encountering repeated deadlocks, because resolving the deadlock for one agent in such a group may lead to others getting stuck again.
\\

If no deadlock is detected in the entire set of agents, the previously generated execution plan is executed, and the process moves on to the next detection loop.

\subsection{Deadlock Resolution Using LLM Strategies with PIBT}

When a deadlock is detected, the LLM analysis is turned into strategies for resolving it. These strategies set the agents’ priorities and the priorities of their actions. The priorities are then applied to PIBT, which create collision-free 1-step actions for each agent. This process follows a loop: after each set of 1-step actions is executed, the strategy is reapplied to PIBT to generate the next set of 1-step actions. This loop continues for the length of the execution plan, progressively resolving the deadlock. The following explains how the strategy forms the priority order of agents and their actions. The order presented below reflects the priority ranking of agents:  \\

\begin{enumerate}
    \item \textbf{Lead agents} (from groups assigned “leader") are the agents farthest from their goals within their respective groups and prioritize moving toward their goals using the shortest path heuristic.
    \item \textbf{Radiate agents} (from groups assigned “radiation") move away from the deadlock center, which is calculated as the average position of all agents in the group, to prevent further deadlock.
    \item \textbf{Non-deadlock agents} prioritize the 1-step actions generated by the base model. While not depicted in Figure \ref{fig:method}, they act independently to ensure the overall efficiency.
    \item \textbf{Yield agents} (from groups assigned “leader") are the remaining agents within each group who are not selected as the lead agent, and they prioritize staying to allow the lead agent to move.
\end{enumerate}

\vspace{\baselineskip}

\begin{algorithm}
\scriptsize
\caption{Strategized PIBT for Deadlock Resolution}
\label{alg:strategizedPIBT}
\raggedright
\textbf{Parameters:} Current states $s^{1:N}$, Actions $a^{1:N}_{1:5}$, Static Obstacles $W$, Strategy $S$ \\
\textbf{Output:} Collision-free strategized actions $a^{1:N}$
\begin{algorithmic}[1]
\Procedure{StrategizedPIBT}{Current states $s^{1:N}$, Actions $a^{1:N}_{1:5}$, Strategy $S$}
    \State agentPriorities $\gets \text{empty list}$
    \State actionsSorted $\gets \text{empty list}$
    \For{each agent $k$}
        \State agentPriorities[k] $\gets S.\text{priorityForAgent}(k)$ \Comment{Get agent priority from strategy $S$}
        \State actionsSorted[k] $\gets \text{sort}(a^k_{1:5}, \text{according to strategy } S.\text{actionPriorityForAgent}(k))$
    \EndFor
    \State collisionFreeActions $\gets$ \textbf{PIBT}($s^{1:N}$, actionsSorted, agentPriorities)
    \State \Return collisionFreeActions
\EndProcedure
\end{algorithmic}
\end{algorithm}

The overall strategy is applied to Algorithm \ref{alg:strategizedPIBT}, where PIBT is used to derive collision-free 1-step action set.After executing these actions, the process loops back to reapply the strategy and generate the next set of actions. This cycle repeats for the execution plan length, aiming to resolve the deadlocks before the next detection loop. Once the cycle is completed for the length of the execution plan, the process moves to the next detection loop, regardless of whether deadlocks remain.


\section{Experimental Results and Analysis}
We conducted three experiments to evaluate LLMDR: (1) comparing its performance with the baseline models in different maps and agent numbers; second, (2) analyzing the impact of the language model on the performance of LLMDR and (3) examining the effect of hyperparameters on LLMDR.

\begin{table*}[htbp]
\centering
\caption{Performance comparison for different Neural MAPF solvers on the warehouse map. We report average episode length (EL, lower is better $\downarrow$) and success rate (SR, higher is better $\uparrow$). EPH* denotes the base model of EPH without its original inference strategies and ensemble.}
\label{tab:warehouse-neural-solvers}
\resizebox{\textwidth}{!}{%
\begin{tabular}{@{}cccccccccccccccccc@{}}
\toprule
\multirow{2}{*}{Map} & \multirow{2}{*}{Agents} & \multicolumn{2}{c}{PRIMAL} & \multicolumn{2}{c}{DHC} & \multicolumn{2}{c}{DHC+LLMDR} & \multicolumn{2}{c}{DCC} & \multicolumn{2}{c}{DCC+LLMDR} & \multicolumn{2}{c}{SACHA} & \multicolumn{2}{c}{EPH*} & \multicolumn{2}{c}{EPH*+LLMDR} \\
\cmidrule(lr){3-4} \cmidrule(lr){5-6} \cmidrule(lr){7-8} \cmidrule(lr){9-10} \cmidrule(lr){11-12} \cmidrule(lr){13-14} \cmidrule(lr){15-16} \cmidrule(l){17-18}
 &  & EL & SR & EL & SR & EL & SR & EL & SR & EL & SR & EL & SR & EL & SR & EL & SR \\
\midrule
\multirow{5}{*}{\texttt{warehouse}} 
 & 4 & 355.80 & 42\% & 146.12 & 99\% & 142.32 & \textbf{100\%} & 135.89 & 99\% & 134.21 & \textbf{100\%} & 134.59 & 99\% & \textbf{133.76} & \textbf{100\%} & 133.9 & \textbf{100\%} \\
 & 8 & 451.82 & 18\% & 198.82 & 91\% & 174.19 & \textbf{99\%} & 169.50 & 96\% & \textbf{160.07} & \textbf{99\%} & 166.72 & 93\% & 173.05 & 95\% & 162.20 & \textbf{99\%} \\
 & 16 & 492.04 & 8\% & 281.37 & 74\% & 206.35 & \textbf{98\%} & 208.72 & 90\% & \textbf{190.58} & 96\% & 198.72 & 76\% & 240.44 & 81\% & 194.42 & 96\% \\
 & 32 & 505.58 & 4\% & 432.28 & 28\% & 258.18 & \textbf{94\%} & 335.81 & 58\% & \textbf{231.50} & 93\% & 354.33 & 48\% & 352.95 & 52\% & 248.18 & 90\% \\
 & 64 & 512.00 & 0\% & 512.00 & 1\% & 350.21 & \textbf{83\%} & 473.92 & 14\% & \textbf{319.02} & 74\% & 437.29 & 28\% & 496.22 & 6\% & 357.72 & 68\% \\
\bottomrule
\end{tabular}%
}
\end{table*}

\subsection{Experiment Setup}
Our experiments evaluate LLMDR's performance across various scenarios, comparing it with baseline models. We conducted tests on four standard MAPF benchmark maps \cite{stern2019multi}: \texttt{random32} and \texttt{random64} are square maps of size (32x32) and (64x64), \texttt{den312d} is a (65x81) game map, and \texttt{warehouse} is a (161x63) warehouse map. For each map, we varied the number of agents, testing with 4, 8, 16, 32, and 64 agents. We set maximum time step to 256 for \texttt{random32}, \texttt{random64}, and \texttt{den312d}, and 512 for the \texttt{warehouse} map. The first experiment included 300 test instances, while the second and the third had 100 each. All base models use a shortest-path heuristic. LLMDR retains this when applied.

We used success rate and average episode length as metrics. Success rate is calculated as the proportion of instances resolved within the maximum time step, while average episode length is the mean number of time steps taken to resolve test instances. For unresolved instances, we used the maximum time step in the average calculation.

In our experiments, we combined LLMDR with the learned models of several learning-based methods, including DCC \cite{ma2021dcc}, DHC \cite{ma2021dhc}, and EPH \cite{tang2024ensembling}. Notably, EPH was used only as a base model without its original inference strategies. We tested with \texttt{gpt-3.5-turbo} and \texttt{gpt-4o} as the underlying language models. Unless stated otherwise, and excluding hyperparameter analysis, we set the detection window length to 4 and the execution plan length to 16 for all tests.

\subsection{LLMDR Performance with Various Base Models}

LLMDR improves the performance of all base models by resolving deadlocks. By improving performance across different models, LLMDR shows potential for broader application in learning-based MAPF methods. Table~\ref{tab:warehouse-neural-solvers} shows how LLMDR improves performance when combined with various base models. In the warehouse map, which is prone to deadlocks, LLMDR consistently improves performance, especially as the number of agents increases. Notably, DHC struggled in the 64-agent scenario, but with LLMDR, it achieved the highest success rate. Similar improvements were observed with other base models. Additionally, increasing the maximum time step could impact the average episode length, affecting performance metrics. These findings suggest that many failures in learned MAPF models stem from deadlocks, and resolving them with LLMDR improves performance.

\subsection{Impact of Language Model Capability on LLMDR}

\begin{table}[htbp]
\centering
\caption{Comparison of DCC and its LLMDR variants using different language models on various maps and agent numbers.}
\label{tab:mapf-solvers-comparison}
\resizebox{\columnwidth}{!}{%
\begin{tabular}{@{}c c cccccc@{}}
\toprule
\multirow{2}{*}{{Map}} & \multirow{2}{*}{Agents} & \multicolumn{2}{c}{DCC} & \multicolumn{2}{c}{DCC+LLMDR} & \multicolumn{2}{c}{DCC+LLMDR} \\
 & & & & \multicolumn{2}{c}{(\texttt{gpt-3.5-turbo})} & \multicolumn{2}{c}{(\texttt{gpt-4o})} \\
\cmidrule(lr){3-4} \cmidrule(lr){5-6} \cmidrule(l){7-8}
 & & EL & SR & EL & SR & EL & SR \\
\midrule
\multirow{5}{*}{\rotatebox[origin=c]{90}{\texttt{random32}}} 
 & 4  & \textbf{33.20}  & \textbf{100\%} & 41.29 & \textbf{100\%} & \textbf{33.2} & \textbf{100\%} \\
 & 8  & 38.63  & \textbf{100\%} & 51.17 & \textbf{100\%} & \textbf{38.23} & \textbf{100\%} \\
 & 16 & \textbf{44.35}  & \textbf{100\%} & 58.09 & \textbf{100\%} & 46.08 & \textbf{100\%} \\
 & 32 & \textbf{51.74}  & \textbf{100\%} & 59.97 & \textbf{100\%} & 53.84 & \textbf{100\%}  \\
 & 64 & 95.67  & 90\%  & 76.22 & \textbf{100\%}  & \textbf{69.52} & \textbf{100\%} \\
\midrule
\multirow{5}{*}{\rotatebox[origin=c]{90}{\texttt{random64}}}
 & 4  & 69.77  & 99\%  & 77.1 & \textbf{100\%} & \textbf{68.08} & \textbf{100\%}  \\
 & 8  & 83.02  & 97\%  & 89.47 & \textbf{100\%} & \textbf{79.17} & \textbf{100\%} \\
 & 16 & 92.64  & 95\%  & 100.59 & \textbf{100\%} & \textbf{85.44} & \textbf{100\%} \\
 & 32 & 116.69 & 88\%  & 111.83 & \textbf{100\%} & \textbf{108.09} & 97\% \\
 & 64 & 147.31 & 77\%  & 138.01 & \textbf{99\%}  & \textbf{132.31} & 94\% \\
\midrule
\multirow{5}{*}{\rotatebox[origin=c]{90}{\texttt{den312d}}}
 & 4  & 85.41  & 99\%  & 94.14 & \textbf{100\%} & \textbf{84.23} & \textbf{100\%} \\
 & 8  & 97.78  & 99\%  & 114.43 & \textbf{100\%} & \textbf{96.57} & \textbf{100\%} \\
 & 16 & \textbf{106.95} & 99\%  & 121.36 & \textbf{100\%} & 106.98 & \textbf{100\%} \\
 & 32 & 124.00 & 94\%  & 131.91 & \textbf{100\%} & \textbf{119.54} &\textbf{100\%}  \\
 & 64 & 149.10 & 90\%  & 146.82 & \textbf{100\%}  & \textbf{141.71} & \textbf{100\%} \\
\midrule
\multirow{5}{*}{\rotatebox[origin=c]{90}{\texttt{warehouse}}}
 & 4  & \textbf{131.16} & \textbf{100\%} & 144.2 & \textbf{100\%} & 131.17 & \textbf{100\%} \\
 & 8  & 179.68 & 93\%  & 175.29 & \textbf{99\%} & \textbf{159.85} & \textbf{99\%}  \\
 & 16 & 212.92 & 89\%  & 198.69 & \textbf{99\%}  & \textbf{184.69} & 98\%  \\
 & 32 & 321.81 & 61\%  & 249.78 & 90\%  & \textbf{228.57} & \textbf{92\%}  \\
 & 64 & 468.48 & 16\%  & 347.96 & 71\%  & \textbf{324.55} & \textbf{74\%}  \\
\bottomrule
\end{tabular}%
}
\end{table}

Our experiments, as shown in Table~\ref{tab:mapf-solvers-comparison}, show that the language model's capability affects LLMDR's performance. Both \texttt{gpt-3.5-turbo} and \texttt{gpt-4o} versions of LLMDR outperformed the base DCC in success rate, but \texttt{gpt-4o} achieved shorter average episode lengths, indicating more efficient path planning. Additionally, while \texttt{gpt-3.5-turbo} tends to exhibit longer average episode lengths than the base model at similar success rates, this effect is much less noticeable in \texttt{gpt-4o}. These results suggest that as the language model’s reasoning improves, LLMDR better distinguishes deadlocks from non-deadlocks, leading to more precise interventions.

\subsection{Hyperparameter Analysis}

\begin{table}[htbp]
\centering
\caption{Hyperparameter analysis on detection window length (DWL) and execution plan length (EPL) for LLMDR.}
\label{tab:warehouse-ablation}
\begin{tabular}{@{}ccccc@{}}
\toprule
DWL & EPL & SR (\%) & EL \\
\midrule
2 & 16 & 71.00 & 323.91 \\
4 & 16 & 74.00 & 324.55 \\
8 & 16 & 73.00 & 325.15 \\
\midrule
4 & 8  & 70.00 & 323.66 \\
4 & 16 & 74.00 & 324.55 \\
4 & 32 & 75.00 & 339.06 \\
\bottomrule
\end{tabular}
\end{table}
We conducted a hyperparameter analysis using the DCC+LLMDR model to evaluate the impact of detection window length and execution plan length. Table~\ref{tab:warehouse-ablation} presents the results, focusing on the scenario of 64 agents in the warehouse map. For the detection window length, we tested values of 2, 4. The results show that increasing the detection window length beyond 4 does not yield significant performance improvements. This suggests that most deadlocks in our scenarios are due to stagnation rather than oscillation or wandering. For the execution plan length, we tested values of 8, 16, and 32. It was observed that a length of 16 offers a balance between providing enough intervention steps to resolve deadlocks and minimizing the risk of decreased efficiency due to unnecessarily prolonged intervention periods.


\section{Conclusion}

In this study, we propose LLMDR, a novel approach that integrates the LLMs to resolve deadlocks and improve scalability of learned MAPF models. LLMDR analyzes plans from base model, using LLMs to detect deadlocks. If a deadlock is detected, LLMDR proposes solutions through LLM inference, devising a strategy to adjust the agents’ actions and priorities, and integrates these adjustments with PIBT algorithm to resolve the deadlocks. Experimental results show that LLMDR significantly improves the performance of several learned models in complex environments. These results highlight LLMDR’s capability to generalize across various models, while also improving scalability. However, the computational cost of LLMs remains a challenge, limiting LLMDR’s use in resource-constrained or real-time scenarios. Future work could explore more efficient LLM architectures or hybrid approaches to reduce computational overhead.


\bibliographystyle{IEEEtran}
\bibliography{sample}

\begin{thebibliography}{10}
\providecommand{\url}[1]{#1}
\csname url@rmstyle\endcsname
\providecommand{\newblock}{\relax}
\providecommand{\bibinfo}[2]{#2}
\providecommand\BIBentrySTDinterwordspacing{\spaceskip=0pt\relax}
\providecommand\BIBentryALTinterwordstretchfactor{4}
\providecommand\BIBentryALTinterwordspacing{\spaceskip=\fontdimen2\font plus
\BIBentryALTinterwordstretchfactor\fontdimen3\font minus \fontdimen4\font\relax}
\providecommand\BIBforeignlanguage[2]{{%
\expandafter\ifx\csname l@#1\endcsname\relax
\typeout{** WARNING: IEEEtran.bst: No hyphenation pattern has been}%
\typeout{** loaded for the language `#1'. Using the pattern for}%
\typeout{** the default language instead.}%
\else
\language=\csname l@#1\endcsname
\fi
#2}}

\bibitem{stern2019multi}
R.~Stern, N.~Sturtevant, A.~Felner, S.~Koenig, H.~Ma, T.~Walker, J.~Li, D.~Atzmon, L.~Cohen, T.~Kumar, \emph{et~al.}, ``Multi-agent pathfinding: Definitions, variants, and benchmarks,'' in \emph{Proceedings of the International Symposium on Combinatorial Search}, vol.~10, no.~1, 2019, pp. 151--158.

\bibitem{li2023intersection}
J.~Li, E.~Lin, H.~L. Vu, S.~Koenig, \emph{et~al.}, ``Intersection coordination with priority-based search for autonomous vehicles,'' in \emph{Proceedings of the AAAI Conference on Artificial Intelligence}, vol.~37, no.~10, 2023, pp. 11\,578--11\,585.

\bibitem{kaduri2020algorithm}
O.~Kaduri, E.~Boyarski, and R.~Stern, ``Algorithm selection for optimal multi-agent pathfinding,'' in \emph{Proceedings of the international conference on automated planning and scheduling}, vol.~30, 2020, pp. 161--165.

\bibitem{ma2017feasibility}
H.~Ma, J.~Yang, L.~Cohen, T.~Kumar, and S.~Koenig, ``Feasibility study: Moving non-homogeneous teams in congested video game environments,'' in \emph{Proceedings of the AAAI Conference on Artificial Intelligence and Interactive Digital Entertainment}, vol.~13, no.~1, 2017, pp. 270--272.

\bibitem{kraus1997negotiation}
S.~Kraus, ``Negotiation and cooperation in multi-agent environments,'' \emph{Artificial intelligence}, vol.~94, no. 1-2, pp. 79--97, 1997.

\bibitem{boyarski2015icbs}
E.~Boyarski, A.~Felner, R.~Stern, G.~Sharon, O.~Betzalel, D.~Tolpin, and E.~Shimony, ``Icbs: The improved conflict-based search algorithm for multi-agent pathfinding,'' in \emph{Proceedings of the International Symposium on Combinatorial Search}, vol.~6, no.~1, 2015, pp. 223--225.

\bibitem{sartoretti2019primal}
G.~Sartoretti, J.~Kerr, Y.~Shi, G.~Wagner, T.~S. Kumar, S.~Koenig, and H.~Choset, ``Primal: Pathfinding via reinforcement and imitation multi-agent learning,'' \emph{IEEE Robotics and Automation Letters}, vol.~4, no.~3, pp. 2378--2385, 2019.

\bibitem{ma2021dhc}
Z.~Ma, Y.~Luo, and H.~Ma, ``Distributed heuristic multi-agent path finding with communication,'' in \emph{2021 IEEE International Conference on Robotics and Automation (ICRA)}.\hskip 1em plus 0.5em minus 0.4em\relax IEEE, 2021, pp. 8699--8705.

\bibitem{ma2021dcc}
Z.~Ma, Y.~Luo, and J.~Pan, ``Learning selective communication for multi-agent path finding,'' \emph{IEEE Robotics and Automation Letters}, vol.~7, no.~2, pp. 1455--1462, 2021.

\bibitem{flammini2024preventing}
B.~Flammini, D.~Azzalini, and F.~Amigoni, ``Preventing deadlocks for multi-agent pickup and delivery in dynamic environments,'' in \emph{Proceedings of the 23rd International Conference on Autonomous Agents and Multiagent Systems}, 2024, pp. 580--588.

\bibitem{okumura2022priority}
K.~Okumura, M.~Machida, X.~D{\'e}fago, and Y.~Tamura, ``Priority inheritance with backtracking for iterative multi-agent path finding,'' \emph{Artificial Intelligence}, vol. 310, p. 103752, 2022.

\bibitem{ferner2013odrm}
C.~Ferner, G.~Wagner, and H.~Choset, ``Odrm* optimal multirobot path planning in low dimensional search spaces,'' in \emph{2013 IEEE international conference on robotics and automation}.\hskip 1em plus 0.5em minus 0.4em\relax IEEE, 2013, pp. 3854--3859.

\bibitem{grenouilleau2019multi}
F.~Grenouilleau, W.-J. Van~Hoeve, and J.~N. Hooker, ``A multi-label a* algorithm for multi-agent pathfinding,'' in \emph{Proceedings of the international conference on automated planning and scheduling}, vol.~29, 2019, pp. 181--185.

\bibitem{liu2020mapper}
Z.~Liu, B.~Chen, H.~Zhou, G.~Koushik, M.~Hebert, and D.~Zhao, ``Mapper: Multi-agent path planning with evolutionary reinforcement learning in mixed dynamic environments,'' in \emph{2020 IEEE/RSJ International Conference on Intelligent Robots and Systems (IROS)}.\hskip 1em plus 0.5em minus 0.4em\relax IEEE, 2020, pp. 11\,748--11\,754.

\bibitem{wang2020mobile}
B.~Wang, Z.~Liu, Q.~Li, and A.~Prorok, ``Mobile robot path planning in dynamic environments through globally guided reinforcement learning,'' \emph{IEEE Robotics and Automation Letters}, vol.~5, no.~4, pp. 6932--6939, 2020.

\bibitem{li2020graph}
Q.~Li, F.~Gama, A.~Ribeiro, and A.~Prorok, ``Graph neural networks for decentralized multi-robot path planning,'' in \emph{2020 IEEE/RSJ international conference on intelligent robots and systems (IROS)}.\hskip 1em plus 0.5em minus 0.4em\relax IEEE, 2020, pp. 11\,785--11\,792.

\bibitem{lin2023sacha}
Q.~Lin and H.~Ma, ``Sacha: Soft actor-critic with heuristic-based attention for partially observable multi-agent path finding,'' \emph{IEEE Robotics and Automation Letters}, 2023.

\bibitem{okumura2023lacam}
K.~Okumura, ``Lacam: Search-based algorithm for quick multi-agent pathfinding,'' in \emph{Proceedings of the AAAI Conference on Artificial Intelligence}, vol.~37, no.~10, 2023, pp. 11\,655--11\,662.

\bibitem{veerapaneni2024improving}
R.~Veerapaneni, Q.~Wang, K.~Ren, A.~Jakobsson, J.~Li, and M.~Likhachev, ``Improving learnt local mapf policies with heuristic search,'' in \emph{Proceedings of the International Conference on Automated Planning and Scheduling}, vol.~34, 2024, pp. 597--606.

\bibitem{tang2024ensembling}
H.~Tang, F.~Berto, and J.~Park, ``Ensembling prioritized hybrid policies for multi-agent pathfinding,'' \emph{arXiv preprint arXiv:2403.07559}, 2024.

\bibitem{gao2023rde}
J.~Gao, Y.~Li, X.~Yang, and M.~Tan, ``Rde: A hybrid policy framework for multi-agent path finding problem,'' \emph{arXiv preprint arXiv:2311.01728}, 2023.

\bibitem{huang2022language}
W.~Huang, P.~Abbeel, D.~Pathak, and I.~Mordatch, ``Language models as zero-shot planners: Extracting actionable knowledge for embodied agents,'' in \emph{International conference on machine learning}.\hskip 1em plus 0.5em minus 0.4em\relax PMLR, 2022, pp. 9118--9147.

\bibitem{kannan2023smart}
S.~S. Kannan, V.~L. Venkatesh, and B.-C. Min, ``Smart-llm: Smart multi-agent robot task planning using large language models,'' \emph{arXiv preprint arXiv:2309.10062}, 2023.

\bibitem{shukla2023lgts}
Y.~Shukla, W.~Gao, V.~Sarathy, A.~Velasquez, R.~Wright, and J.~Sinapov, ``Lgts: Dynamic task sampling using llm-generated sub-goals for reinforcement learning agents,'' \emph{arXiv preprint arXiv:2310.09454}, 2023.

\bibitem{zhu2023ghost}
X.~Zhu, Y.~Chen, H.~Tian, C.~Tao, W.~Su, C.~Yang, G.~Huang, B.~Li, L.~Lu, X.~Wang, \emph{et~al.}, ``Ghost in the minecraft: Generally capable agents for open-world environments via large language models with text-based knowledge and memory,'' \emph{arXiv preprint arXiv:2305.17144}, 2023.

\bibitem{garg2024large}
K.~Garg, J.~Arkin, S.~Zhang, N.~Roy, and C.~Fan, ``Large language models to the rescue: Deadlock resolution in multi-robot systems,'' \emph{arXiv preprint arXiv:2404.06413}, 2024.

\bibitem{chen2024solving}
W.~Chen, S.~Koenig, and B.~Dilkina, ``Why solving multi-agent path finding with large language model has not succeeded yet,'' \emph{arXiv preprint arXiv:2401.03630}, 2024.

\bibitem{Erdmann-1987-15683}
M.~Erdmann and T.~Lozano-Perez, ``On multiple moving objects,'' \emph{Algorithmica}, vol.~2, pp. 477 -- 521, January 1987.

\end{thebibliography}

\end{document}